\documentclass[11pt,a4paper]{article}

\usepackage{balance}
\usepackage{comment}
\usepackage{graphicx}
\usepackage{ifthen}
\usepackage{multirow}
\usepackage{paralist}
\usepackage{subfig}
\usepackage{url}
\usepackage{xcolor}

\newcommand{\subfigurewidth}{0.24\textwidth}
\newcommand{\vspacecut}{0cm}

\providecommand{\e}[1]{\ensuremath{\times 10^{#1}}}

\begin{document}

\title{The Effect of Network and Infrastructural Variables on SPDY's Performance}

\author{
  Yehia Elkhatib\\
  The School of Computing \& Communications, Lancaster University\\
  \texttt{y.elkhatib@lancaster.ac.uk}
  \and
  Gareth Tyson\\
  EECS, Queen Mary, University of London\\
  \texttt{gareth.tyson@eecs.qmul.ac.uk}
  \and
  Michael Welzl\\
  Department of Informatics, University of Oslo\\
  \texttt{michawe@ifi.uio.no}
}

%\today
\maketitle

%%%%%%%%%%%%%%%%%%%%%%%%%%%%%%%%%%%%%%%%%%%%%%
\begin{abstract}
HTTP is a successful Internet technology on top of which a lot of the web resides. However, limitations with its current specification, i.e.\ HTTP/1.1, have encouraged some to look for the next generation of HTTP. In SPDY, Google has come up with such a proposal that has growing community acceptance, especially after being adopted by the IETF HTTPbis-WG as the basis for HTTP/2.0. SPDY has the potential to greatly improve web experience with little deployment overhead. However, we still lack an understanding of its true potential in different environments. This paper seeks to resolve these issues, offering a comprehensive evaluation of SPDY's performance using extensive experiments. We identify the impact of network characteristics and website infrastructure on SPDY's potential page loading benefits, finding that these factors are decisive for SPDY and its optimal deployment strategy. Through this, we feed into the wider debate regarding HTTP/2.0, exploring the key aspects that impact the performance of this future protocol.
\end{abstract}

%%%%%%%%%%%%%%%%%%%%%%%%%%%%%%%%%%%%%%%%%%%%%%
%%%%%%%%%%%%%%%%%%%%%%%%%%%%%%%%%%%%%%%%%%%%%%
\section{Introduction}
\label{sec:introduction}

Web pages are constantly increasing in complexity and size. The HTTP Archive reports that the global average web page size has surpassed 1MB in April 2012 \cite{httparchive}. By the start of July 2013, visiting one of the top 1000 sites incurs, on average, 1246kB of web page resources over 100 separate requests \cite{httparchive}. 
Such growth has been fueled by the emergence of advanced web-based services (Web 2.0, SaaS cloud services, etc.), enhanced client device capabilities (JavaScript browser runtimes \cite{Severance2012javascript}, display), and increased downlink speeds \cite{akamai2012q4soti}. 
This growing complexity, however, can dramatically slow down page retrieval. Unfortunately, this has negative consequences \cite{skadberg2004visitors}, 
and very real ones in the case of commercial websites. It has been found that most users cannot tolerate in excess of 2 seconds of page load delay \cite{nah2004study}, and that increments of just 100ms on shopping websites can decrease sales by 1\% \cite{Kohavi2007experiments}. The converse is similarly true: 
decreasing delay can have a powerful enhancing effect, with Google claiming to have increased ad revenue by 20\% through cutting 500ms from load times. To mitigate page load times, various extensions to HTTP have been proposed. 
However, in practice, little progress has been made, with many web servers, proxies and browsers being slow to adopt these new tweaks (e.g.\ pipelining \cite{rfc2616}).

In light of these observations, some have proposed developing a new web protocol. Such efforts include Microsoft's Speed+Mobility \cite{SM-draft} and HTML5 Websockets; most prominent, however, is Google's SPDY \cite{SPDY-draft}. This has already begun to see deployment by prominent organisations such as Google, Twitter, Akamai and Facebook, whilst also being adopted as the base for HTTP/2.0 by the HTTPbis Working Group. Despite this, we still possess a limited understaning of its behaviour, overheads and performance: does it offer a fundamental improvement or just further tweaking?

In an attempt to answer the above question, a small number of early stage studies have explored the topic. 
They offer a range of results, with some claiming significant gains and (curiously) others claiming rather negative results. This report seeks to resolve these issues, by analysing the circumstances under which SPDY can improve page load times, and the ones where the opposite is true. 

To achieve this, we perform a large-scale evaluation of SPDY using an emulated testbed. 
Based on NPN negotiation handshakes, we found out that the number of top 10,000 Alexa sites adopting SPDY was 208 on October 14\textsuperscript{th} 2012, rising to 271 on April 23\textsuperscript{rd} 2013. 
Using some of these websites, we execute a large number of probes to measure the performance of SPDY in the wild. Confirming our suspicions, we find highly variable results between different websites and samples: SPDY has the potential to both benefit and damage page load times. Motivated by this, we perform a large body of controlled experiments in our local testbed to understand the reasons behind these performance variations. We identify the website types and network characteristics that SPDY thrives under, as well as how these benefits vary based on provider-side infrastructural decisions. To our knowledge, this is the first effort to offer such insights.

The rest of the report is organized as follows. Section \ref{sec:background} provides background and highlights related work. Section \ref{sec:meth} describes our measurement toolkit and environments. Section \ref{sec:results:live} presents the results of comparing SPDY to HTTPS on live websites. We then use an emulated network testbed in order to dissect the factors affecting SPDY performance, namely network characteristics (section \ref{sec:results:network}) and infrastructure setup (section \ref{sec:results:infra}). 
Section \ref{sec:Conclusion} concludes and discusses future work.

%%%%%%%%%%%%%%%%%%%%%%%%%%%%%%%%%%%%%%%%%%%%%%
%%%%%%%%%%%%%%%%%%%%%%%%%%%%%%%%%%%%%%%%%%%%%%
\section{Background and Related Work}
\label{sec:background}

\subsection{SPDY} \label{sec:background:spdy}

SPDY is an application-layer web protocol that reuses HTTP's semantics \cite{rfc2616}. As such, it retains all features including cookies, ETags and Content-Encoding negotiations. SPDY only replaces the manner in which data is written to the network. The purpose of this is to reduce page load time. It does this by introducing the following mechanisms:

\begin{itemize}
 \item \emph{Multiplexing}: A framing layer multiplexes streams over a single connection, removing the need to establish separate TCP connections for transferring different page resources.
 \item \emph{Compression}: All header data is compressed to reduce the overheads of multiple related requests. 
 \item \emph{Universal encryption}: SPDY is negotiated over SSL/TLS and thus operates exclusively over a secure channel in order to address the increasing amounts of traffic sent over insecure paths (e.g.\ public WiFi).
 \item \emph{Server Push/Hint}: Servers could proactively \textit{push} resources to clients (e.g.\ scripts and images that will be required). Alternatively, SPDY can send \textit{hints} advising clients to pre-fetch content.
 \item \emph{Content prioritization}: A client can specify the preferred order in which resources should be transferred.
\end{itemize}

SPDY consists of two components. The first provides framing of data, thereby allowing things like compression and multiplexing. The framing layer works on top of secure (SSL/TLS) persistent TCP connections that are kept alive as long as the corresponding web pages are open. Clients and servers exchange \emph{control} and \emph{data} frames, both of which contain an 8 bytes header. Control frames are used for carrying connection management signals and configuration options, while data frames carry HTTP requests and responses. The second component maps HTTP communication into SPDY data frames. Multiple logical HTTP streams can be multiplexed using interleaved data frames over a single TCP connection. 

\subsection{Related Studies}

There have been a small number of preliminary studies looking at the performance of SPDY. 
The first was presented in a Google white paper \cite{SPDYWhitepaper}. By running 25 of the top 100 websites over simulated home networks, this publication showed significant performance benefits over both HTTP (27-60\%) and HTTPS (39-55\%). Whilst this does seem impressive, there are somewhat conflicting accounts of SPDY's performance in studies provided by Akamai \cite{NotAsSPDY} and Microsoft \cite{Padhye2012TR}. Tests performed by Akamai showed only a marginal benefit over HTTPS, alongside a decrease in performance when compared to HTTP. It was found that SPDY, on average, was only about 4.5\% faster than HTTPS, and about 3.4\% slower than HTTP. Microsoft offered slightly more positive results, but still did not attain the high levels previously reported by Google. 
An Internet Draft by White \emph{et al.} \cite{whiteIEFT} also found a mix of results, highlighting that SPDY's performance would likely depend on a number of factors, e.g. number of servers, TCP configurations, network characteristics (specifically, latency and loss). 
The authors found an average improvement of 29\% over HTTPS. They also reported that such benefits are not necessarily universal (i.e. they vary for different webpages). Nevertheless, the Internet Draft does not offer insight into such findings but rather focuses on coupling SPDY with an increase in the TCP initial congestion window.

It is difficult to derive a direct conclusion from the above studies as they offer a wide range of rather conflicting results. Some claim SPDY outperforms HTTP, whereas others claim the opposite. Consequently, the only clear conclusion is the SPDY has the potential to have highly variable performance. As of yet, these studies do not elucidate this observation, i.e.\ detailing the reasons behind such variations. 

%%%%%%%%%%%%%%%%%%%%%%%%%%%%%%%%%%%%%%%%%%%%%%
%%%%%%%%%%%%%%%%%%%%%%%%%%%%%%%%%%%%%%%%%%%%%%
\section{Measurement Methodology}
\label{sec:meth}

\subsection{Measurement Toolkit}
\label{sec:meth:toolkit}

SPDY is designed to mitigate page load time for end users. We therefore focus on client-side measurements, for which we have built a toolkit based on the Chromium browser. 
This seems a logical choice considering that both SPDY and Chromium were developed by Google. Chromium also offers sophisticated logging features that allow us to extract statistics via automated scripting. We use Chromium\,25 (running over Ubuntu Desktop 12.04.2) via the Chrome-HAR-capturer \cite{chrome-har-capturer} package, which interacts with Chromium through its remote debugging API. To ensure authenticity, we maintained all of Chromium's default settings, apart from disabling DNS pre-fetching in order to include DNS lookup time in all measurements.

When invoked, our measurement toolkit instructs Chromium to fetch a particular webpage. Once this is completed, the toolkit extracts detailed logs in the form of HTTP Archives (HAR) and Wireshark network traces. It then processes them to calculate metrics of interest. Traditionally, page load time has been measured by the Document Object Model (DOM) being fully loaded. 
However, this is not suitable for our purposes, as it also captures browser processing time that is strictly HTML-related, e.g.\ arbitrating the style hierarchy. Instead, we wish to only measure the time spent performing network interactions. Thus, we use an alternate metric which we term the \textit{Time on Wire} (ToW). This is calculated using Wireshark network traces as the period between the first request and last response packets, giving us precise timestamps for the page transmission delay.

Using this toolkit, we employ Chromium in a non-obtrusive manner to retrieve a number of webpages in a range of different environments. The collated measurements allow us to explore the performance of SPDY. The rest of this section details the environments we utilised the measurement toolkit in. 

\subsection{Measurement Setups}
\label{sec:meth:setups}

The measurements are separated into two groups, both using the above toolkit. First, we perform \emph{live tests}, probing real-world deployments (e.g.\ YouTube) to calculate the performance advantages of organisations currently using SPDY. Second, we expand on these results using \emph{emulated network tests}, creating our own controlled SPDY deployment in a local testbed. The latter allows us to deep-dive into SPDY's performance in a deterministic fashion by varying and monitoring the impact of various key factors (namely network conditions and website infrastructure setup). In both cases, a large number of samples are taken to ensure statistical significance. Overall, we have collected over 70,000 probes (12,000 live and 58,000 controlled).

%%%%%%%%%%%%%%%%%%%%%%%%%%%%%%%%%%%%%%%%%%%%%%
\subsubsection{Live Tests}
\label{sec:meth:setups:live}

First, we perform live experiments using web sites that have already deployed SPDY in their real infrastructures.  To discover these, we have implemented a crawler to probe the top 10k Alexa websites\footnote{All Alexa ranks henceforth are of April 23\textsuperscript{rd} 2013.}, recording their individual protocol support. We then select the top 8 Alexa websites that implement SPDY. We choose only the highest Alexa ranked website from every distinct online presence and disregard similar sites (facebook.com (\#1) but not fbcdn.net (\#202); google.com (\#2) but not google.co.in (\#12), google.com.hk (\#22), etc.; and so on). The list is shown in Table \ref{tab:LiveSPDYWebsites}.

\begin{table}[t!]
  \centering
  \caption{Live SPDY-enabled Websites}
  \label{tab:LiveSPDYWebsites}
    \begin{tabular}{l|rrrccc}
                                  & \multicolumn{3}{c}{Resources}     & SPDY              &     & Av.RTT \\ \cline{2-4}
        Site                      & Count   & Av. Size (kB) & Domains & Version           & IW  &  (ms)   \\ \hline

        Facebook                  & 20      & 12.56         & 4       & 2                 & 7   &   92    \\        
        Google                    & 7       & 41.29         & 2       & 3                 & 7   &    8    \\
        YouTube                   & 50      & 10.63         & 4       & 3                 & 7   &    8    \\
        Blogspot                  & 31      & 5.03          & 6       & 3                 & 7   &   17    \\
        Twitter                   & 7       & 46.40         & 3       & 3                 & 10  &  158    \\
        WordPress                 & 13      & 7.92          & 4       & 2                 & 10  &   91    \\
        imgur                     & 133     & 11.78         & 58      & 2                 & 10  &    8    \\
        youm7                     & 270     & 11.07         & 54      & 2                 & 10  &  150    \\
    \end{tabular}
\end{table}

The selected websites provide a range of resource sizes, counts, and domains. 
In terms of their respective delivery infrastructures, we note that all appear to use CDNs with the exception of WordPress. We confirm this using \texttt{whois} as well as other means (e.g.\ trying to directly access a CloudFlare IP address with a browser yields an error message \emph{from} CloudFlare). 
It seems, unsurprisingly, that the employed CDN dictates the supported SPDY version and the TCP Initial Window (IW).\footnote{Note that increasing the IW size is another closely related component in Google's ``Make the Web Faster'' project. We determined IW by sending self-crafted TCP packets to the servers. Our program carried out a successful handshake followed by sending a HTTP GET for a large resource (typically a static image). We then noted the number of ensuing packets (which we never acknowledged) as  the server's IW.} We therefore posit that our results for these sites are representative of the performance for other customers of the same respective CDN.

Using our measurement toolkit, we periodically probe each website from the Lancaster University campus using HTTP, HTTPS and SPDY. In this report, we focus on HTTPS as a baseline comparison as, like SPDY, it encrypts its data. However, where possible, we also include HTTP, considering that many websites have no interest in securing their connections. In both cases, when HTTP and HTTPS are used, we avoid bias by forcing Chromium to pursue the Next Protocol Negotiation (NPN) handshake nonetheless as SPDY does. 
The probes are carried out for each website in an alternating sequence of protocols with 2 seconds between each run. For SPDY, we select the highest non-experimental version that the server supports (listed in Table\,\ref{tab:LiveSPDYWebsites}). Tests were carried out from different sites: Lancaster, Dublin, and Tokyo. We only discuss the Lancaster set as the other results provide very similar outcomes. The Lancaster tests ran on weekdays between 12pm and 5pm BST between 20/5/2013 and 23/5/2013. In total, we performed 1.06 million GET requests, with 500 samples taken for each website and protocol combination.

%%%%%%%%%%%%%%%%%%%%%%%%%%%%%%%%%%%%%%%%%%%%%%
\subsubsection{Emulated Network Tests}
\label{sec:meth:setups:emu-net}

The above live experiments provide useful context to the current state of affairs, but are somewhat limited in what they can tell us. Although the comparison they provide is fair, it would be difficult to definitely and neutrally ascertain SPDY's performance as it is subject to variations in the network, and tightly bound to the particular deployment under test: its characteristics (e.g.\ web server, SPDY module/proxy) and its status (e.g.\ server load). To address these issues, we extend our tests by creating our own SPDY deployment in an emulated testbed interconnected via a LAN. This allows us to control the various network parameters to understand how they impact performance.

Our testbed consists of a client and server setup. The client runs our measurement toolkit and is connected via 100Mbps Ethernet. We then emulate various network conditions: the Linux \texttt{tc} utility is used to throttle bandwidth by shaping traffic with Hierarchy Token Bucket queuing \cite{htb}, and NetEm \cite{hemminger2005netem} is used (at the server) to specify a deterministic round trip time (RTT) and packet loss ratio (PLR).

The server runs Ubuntu Server 12.04 with the Apache 2.2.22 web server, supporting both HTTP and HTTPS. We use Apache's mod\_spdy 0.9.3.3-386 module which implements spdy/3. This is the most advanced SPDY implementation available, provided by Google's own SPDY project \cite{mod-spdy}. Using this server, we clone a set of the SPDY websites discovered in the wild; these are each intended to be representative of a broader class of top Alexa websites and are as follows:

\begin{itemize}
 \item \emph{Twitter}: This is a simple page with only a few (7) resources (average size 46.4KB). This is comparable to other top Alexa websites such as Google and Blogspot and their regional versions, Wikipedia, and Soso (Chinese search engine).
 \item \emph{YouTube}: This is a relatively complicated page, with a fair number (50) of resources (average size 10.63KB).
 Examples of similar websites include Amazon's regional websites, AOL, Alibaba (Chinese e-commerce portal), About.com, and DailyMotion.
 \item \emph{imgur}: This is a complicated page, with a large number (133) of images and flash (average size 11.78KB). Similar websites include QQ (Chinese messaging website), TaoBao (China's equivalent to eBay), The New York Times, CNN, and MSN.
\end{itemize}

The described setup enables us to have a single server-side SPDY implementation and a single SPDY-capable web browser, which rules out software discrepancies (note that we also later use multiple servers). This method also allows us to control the network characteristics in order to experiment with SPDY under different network conditions. Moreover, server and connection load are also controlled. Using this testbed, we apply the same methodology as in the live experiments, generating repeated page requests for the chosen webpages using HTTPS and SPDY.
The exact details of the parameters investigated will be presented in Sections \ref{sec:results:network} and \ref{sec:results:infra}. 

%%%%%%%%%%%%%%%%%%%%%%%%%%%%%%%%%%%%%%%%%%%%%%
%%%%%%%%%%%%%%%%%%%%%%%%%%%%%%%%%%%%%%%%%%%%%%
\section{Live Results}
\label{sec:results:live}

\begin{figure*}[t!]
  \centering
    \subfloat[Facebook]{ \includegraphics[width=\subfigurewidth]{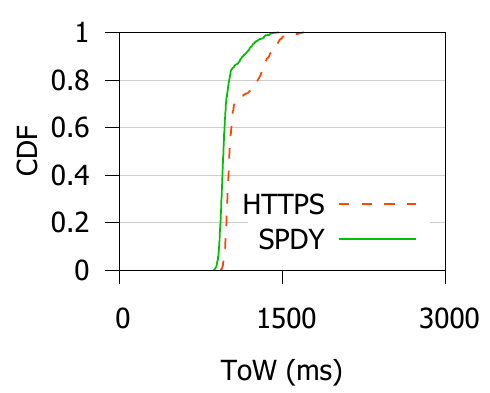} }
    \subfloat[Google]{ \includegraphics[width=\subfigurewidth]{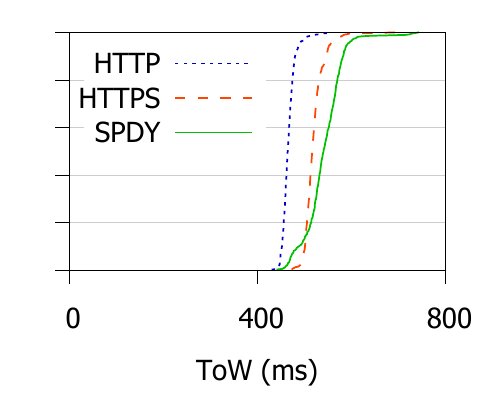} }
    \subfloat[YouTube]{ \includegraphics[width=\subfigurewidth]{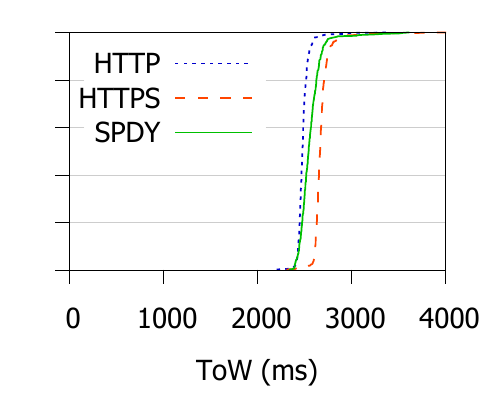} }
    \subfloat[Blogspot]{ \includegraphics[width=\subfigurewidth]{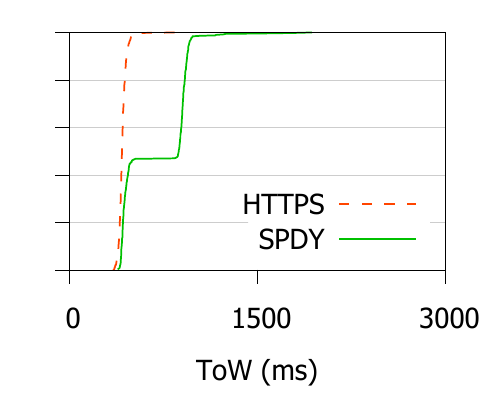} }
    \\
    \subfloat[Twitter]{ \includegraphics[width=\subfigurewidth]{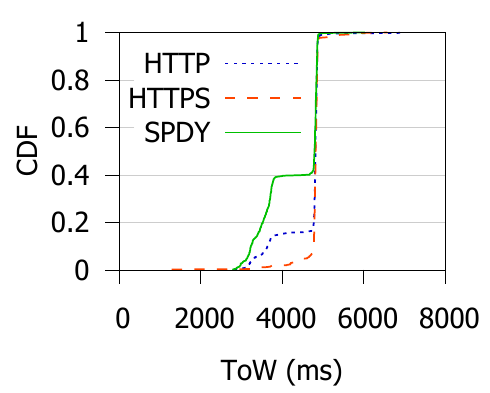} }
    \subfloat[WordPress]{ \includegraphics[width=\subfigurewidth]{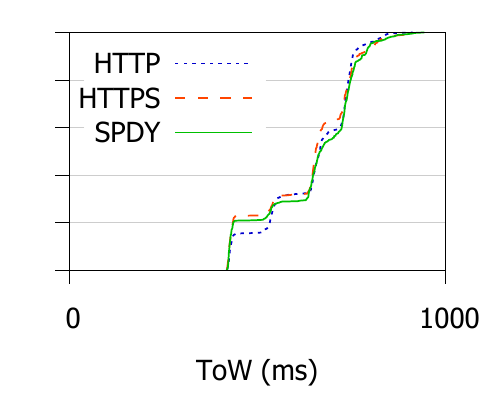} }
    \subfloat[imgur]{ \includegraphics[width=\subfigurewidth]{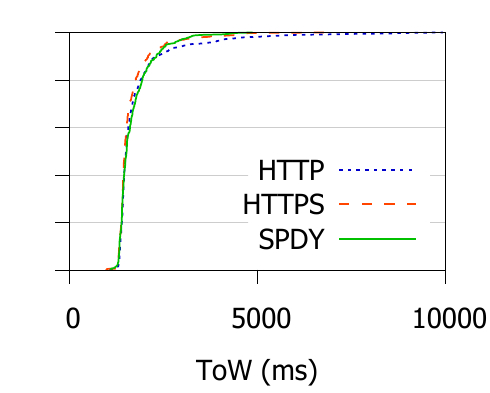} }
    \subfloat[youm7]{ \includegraphics[width=\subfigurewidth]{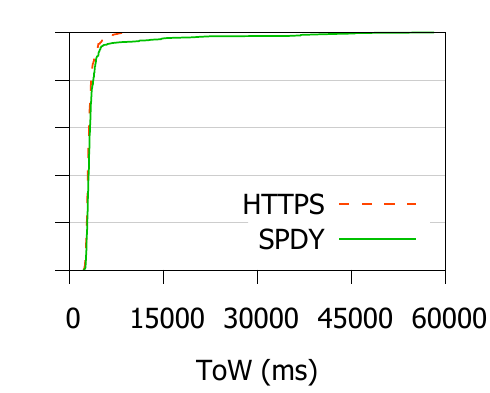} }
  \caption{ToW of Live SPDY-enabled Websites}
  \label{fig:plotLiveTop8-ToW}
\end{figure*}

We begin by performing experiments with existing deployments of SPDY. The aim here is not an exhaustive study but, rather, to form a general idea of the benefits being gained by some of those who have so far adopted SPDY. To achieve this, each website in Table \ref{tab:LiveSPDYWebsites} 
is probed 500 times to calculate its ToW. Figure \ref{fig:plotLiveTop8-ToW} displays the cumulative distribution functions (CDFs) of these measurements, and Table \ref{tab:LiveSPDYgain} summarises them. Note that the HTTP results 
are not included for websites that redirect such requests to HTTPS.  

\vspace{\vspacecut}
\begin{table}[h!]
  \centering
  \caption{Gain in ToW for Live SPDY-enabled Websites}
  \label{tab:LiveSPDYgain}
  \begin{tabular}{l|r}
        & Average Gain in ToW \\
    Site    & (SPDY over HTTPS) \\ \hline
    Facebook  & 7.0\% \\
    Google    & -20.2\% \\
    YouTube   & 4.7\% \\
    Blogspot  & -6.0\% \\
    Twitter   & 10.6\% \\
    WordPress   & -15.1\% \\
    imgur     & 0.8\% \\
    youm7     & 9.7\% \\
  \end{tabular}
\end{table}

Confirming our analysis of past studies, the results are \emph{not} conclusive. We find no clear winner among the three protocols. Instead, we observe large performance variations between different websites, as well as between different samples for the same website. We find that notable improvements are, indeed, gained in some cases. On average, ToW is reduced by 7\% for Facebook, 4.7\% for YouTube, and 9.7\% for youm7. The biggest winner is the Twitter front-page, with an average ToW reduction of 10.6\%. This, however, is not a universal observation. In other cases, improvements are far more modest; for example, imgur only achieves a meager improvement of 0.8\%. Moreover, we find websites that suffer from their use of SPDY; an average ToW increase of 
6.0\% for Blogspot and 15.1\% for Wordpress. Ironically, the biggest sufferer is Google with a 20.2\% 
increase in ToW for their search homepage. There is certainly no one-size-fits-all operation with SPDY, as all websites alternate between SPDY and HTTP optimality.

These experiments therefore raise some interesting (yet serious) questions. From a research perspective, one might ask why these notable variations occur? From an administrator's perspective, the next logical question would then be if SPDY would benefit their deployment? The remainder of this report now explores these questions using emulated experiments. Whereas the live experiments limit our control to the client-side, emulated experiments allow us to dissect all aspects to understand the causes of such variations.

%%%%%%%%%%%%%%%%%%%%%%%%%%%%%%%%%%%%%%%%%%%%%%
\section{Effect of Network Performance}
\label{sec:results:network}
%%%%%%%%%%%%%%%%%%%%%%%%%%%%%%%%%%%%%%%%%%%%%%

To understand the reasons behind the variations in performance witnessed in the wild, we now perform controlled experiments in our testbed. We aim to examine the effect of different network conditions on the performance gain of SPDY over HTTPS. We mirror three of the above representative websites (Twitter, YouTube, imgur) and measure their ToW on a single client, single server testbed under a variety of network characteristics regarding delay, bandwidth, and loss.

%%%%%%%%%%%%%%%%%%%%%%%%%%%%%%%%%%%%%%%%%%%%%%
\subsection{Delay}
\label{sec:results:network:rtt}

\begin{figure}[htb!]
  \centering
  \includegraphics[width=80mm]{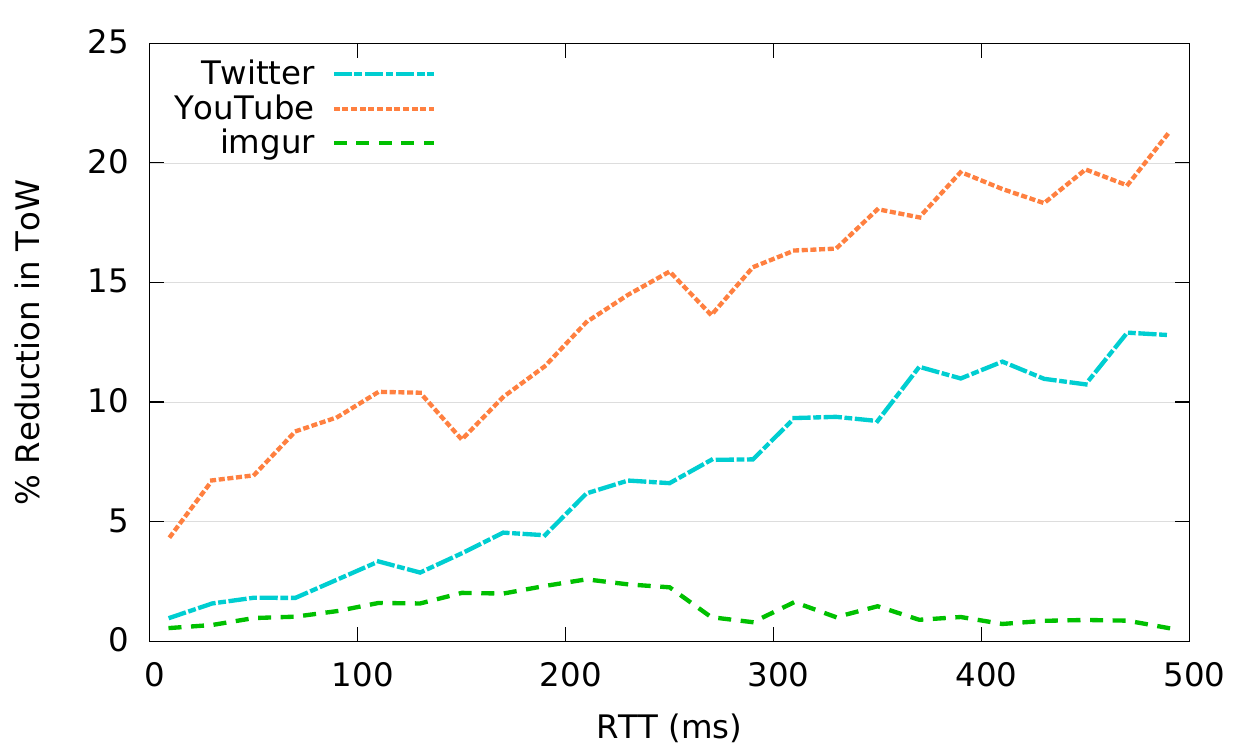}
  \caption{Effect of Round Trip Time (BW=1Mbps, PLR=0\%).}
\label{fig:emu-delay-bw1}
\end{figure}

First, we inspect the impact that round trip time (RTT) has on SPDY's performance. In a real environment, this varies a lot between different requests due to client locations and path characteristics \cite{Kaune2009modelling,Elkhatib11}. To remove any variance, we fix bandwidth (BW) at 1Mbps and Packet Loss Ratio (PLR) at 0\%, whilst changing the RTT between the client and server in a range from 10ms to 490ms. Following this, we perform 20 requests for each website using each configuration with both SPDY and HTTPS. The results are presented in Figure \ref{fig:emu-delay-bw1} as the average percentage improvement in ToW of SPDY over HTTPS. 

In contrast to the live experiments, we see that SPDY \emph{always} achieves better performance than HTTPS in this setup. With low RTTs, these benefits are marginal: requests with RTTs below 150ms achieve under 5\% improvement on average. These benefits, however, increase dramatically as the RTT goes up. In the best case (490ms RTT for YouTube), SPDY beats HTTPS by 21.26\%. 
The results effectively highlight the key benefit of SPDY: stream multiplexing. As RTT goes up, it becomes increasingly expensive for HTTPS to establish separate connections for each resource. \emph{Each} HTTPS connection costs one round trip on TCP handshaking and a further two on negotiating SSL setup. SPDY does this only once (per server) and hence reduces such large waste by multiplexing streams over a single 
connection. By inspecting the HAR logs, we find that SPDY saves between 66\% and 94\% of SSL setup time, creating significant gains in high delay settings.

There is also a notable variation between the different webpages under test. In the cases of Twitter and YouTube, SPDY's ability to multiplex is well exploited by retrieving Twitter's 7 resources and YouTube's 50 resources in parallel. YouTube is by far the greatest beneficiary from SPDY with an average improvement of 13.81\% over HTTPS, whilst Twitter comes second with 6.87\%. The benefits for Twitter are less pronounced because there are fewer streams that can be multiplexed, therefore reducing the benefits of SPDY over HTTPS's maximum of 6 parallel TCP connections (note that this limit of six is hard coded, based on the amendment \cite{httpbisTicket131} to the limit set by RFC 2616 \cite{rfc2616}). 

\begin{figure}[htb!]
  \centering
    \includegraphics[width=80mm]{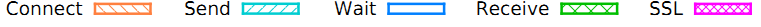} \\
    \includegraphics[width=80mm]{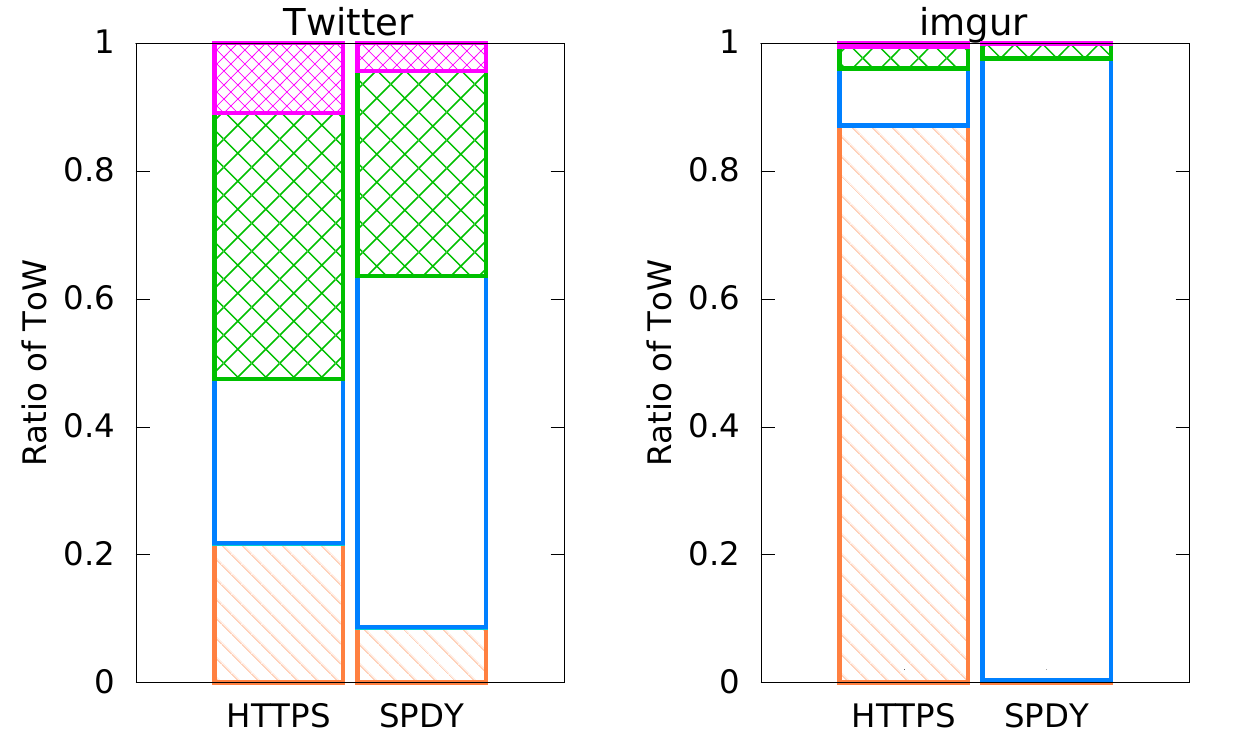}
  \caption{Breakdown of HAR Times for Twitter and imgur at RTT=250. SPDY spends 49\% in Wait for Twitter, and 97\% for imgur due to its high resource count.}
  \label{fig:plot-HARtimes-bw2-rtt490ratios}
\end{figure}

Perhaps more interesting, though, is the fairly steady behaviour exhibited by imgur across all delay values. At first, one would imagine imgur to benefit greatly from SPDY due to its ability to multiplex imgur's large number of resources (133). However, performance is very subdued: its overall average is 1.32\%. To understand this, we inspect the HAR logs to see what is occuring `under the bonnet'. Figure \ref{fig:plot-HARtimes-bw2-rtt490ratios} depicts a breakdown of the HTTPS and SPDY retrieval times for Twitter and imgur at RTT=490ms and BW=2Mbps. We choose this particular subset of our experiments as it provides network conditions where both websites achieve equal SPDY-induced improvement ($\approx$15\%), and hence provides a fair comparison. The figure shows the fraction of time spent in the five key stages of page retrieval: connect, send, wait, receive and SSL. We notice that the make-up of these retrievals is remarkably different. As expected, HTTPS spends a lot of time in the connect and SSL phases, establishing TCP and SSL connections (respectively). This increases for imgur, which has 19 times as many resources as Twitter. On the other hand, SPDY greatly reduces the connect and SSL stages but spends a huge proportion of time in wait. This phase begins when the browser issues a request for a resource, and ends when an intial response is received back. The receive phase is time spent receiving the response data until it is loaded into the browser's memory. In the case of SPDY, wait includes not just the network latency between the client and server but also the time requests are blocked until multiplexed onto the wire. For imgur, SPDY cuts connect time by 94\% but inflates wait time by more than 9 times. As emulated RTT was the same for both protocols, it appears that this inflation in wait time is an unfortunate product of SPDY's multiplexing, but we are unable to exactly ascertain \emph{why} multiplexing is creating this much delay. 
In other words, SPDY's savings in establishing new connections is compromised with multiplexing overhead for highly complex webpages served over a single connection.

\vspace{\vspacecut}

%%%%%%%%%%%%%%%%%%%%%%%%%%%%%%%%%%%%%%%%%%%%%%
\subsection{Bandwidth}
\label{sec:results:network:bw}

\begin{figure}[htb!]
  \centering
  \includegraphics[width=80mm]{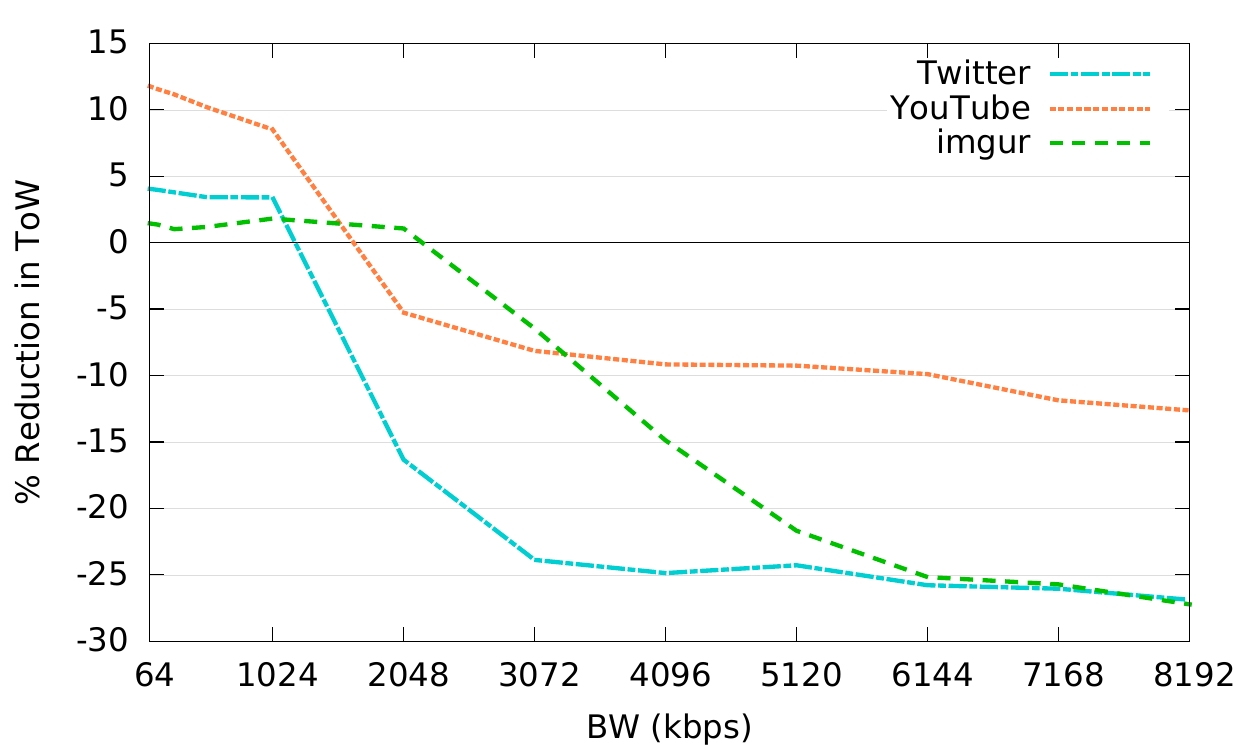}
  \caption{Effect of Bandwidth (RTT=150ms, PLR=0\%).}
\label{fig:emu-bw-rtt150}
\end{figure}

Next, we inspect the impact that client bandwidth has on performance. Once again, this parameter spans a wide range of values across the globe \cite{Dischinger2007broadband,Sundaresan2011broadband,akamai2012q4soti}. 
This time we fix RTT at 150ms and PLR at 0.0\%, while setting client bandwidth to values between 64Kbps and 8Mbps. A first in first out tail-drop queue of 256 packets length is used to emulate commodity routers commonly used as residential and public gateways \cite{Li2007measuring,Gettys2011Bufferbloat}. The results are presented in Figure \ref{fig:emu-bw-rtt150}.

\vspace{\vspacecut}

This graph reveals a very different story to that of delay. Confirming the findings of the live experiments, we see that SPDY \emph{does} have the potential to lower performance, and significantly so. This occurs with a clear trend that favours lower capacities. At 64Kbps, on average, clients witness a 5.75\% improvement over HTTPS, compared to a 22.24\% decrease at 8Mbps. Initial impressions suggest that bandwidth variations have a larger detrimental impact on SPDY's performance.

We now have two dimensions of impact --- RTT and bandwidth --- where SPDY prefers high delay, low bandwidth ($<$1Mbps) environments. As previously discussed, the reason behind SPDY's sensitivity to RTT is relatively easy to measure by inspecting the HAR logs. However, its relationship with bandwidth is rather more complicated. 
To understand this, we turn our attention to the network traces. We find that the separation between RTT and bandwidth is not particularly distinct. This is because HTTPS tends to operate in a somewhat network-unfriendly manner, creating queueing delays where bandwidth is low. The bursty use of HTTPS' parallel connections creates congestion at the gateway queues, 
causing upto 3\% PLR and inflating RTT by upto 570\%\footnote{We also experimented with different gateway queue sizes. Generally, increasing queue size caused longer delays and more loss: upto a 920\% RTT increase and 5\% PLR with a 512 packets queue size, but only 296\% maximum RTT inflation and 1\% PLR with a queue of 64 packets.}. In contrast, SPDY causes negligible packet loss at the gateway.

The network friendly behaviour of SPDY is particularly interesting as Google has recently argued for the use of a larger IW for TCP \cite{rfc6928}. The aim of this is to reduce round trips and speed up delivery --- an idea which has been criticised for potentially causing congestion. One question here is whether or not this is a strategy that is specifically designed to operate in conjunction with SPDY. To explore this, we run further tests using IW=$\left\{3,7,10,16\right\}$ and bandwidth fixed at 1Mbps (all other parameters as above). For HTTPS, it appears that the critics are right: RTT and loss increase greatly with larger IWs. In contrast, SPDY achieves much higher gains when increasing the IW without these negative side effects. It therefore seems that Google have a well integrated approach in their ``Make the Web Faster'' project. 
Interestingly, we observe that the key reason that this increase in RTT and loss adversely affects HTTPS is that it slows down the connection establishment phase, creating a similar situation to that presented earlier in Figure\,\ref{fig:emu-delay-bw1}. Obviously, this congestion also severely damages window ramping over the HTTPS connections. We can tangibly observe this by inspecting the client's TCP window size, which scales far faster with SPDY than any one of the parallel HTTPS connections; this alone leads to an average of $\approx$10\% more throughput than that of HTTPS.

\begin{figure}[tbh!]
  \centering
  \includegraphics[width=80mm]{HAR-key} \\
  \includegraphics[width=90mm, clip=true, trim=0 0 0 0.75cm]{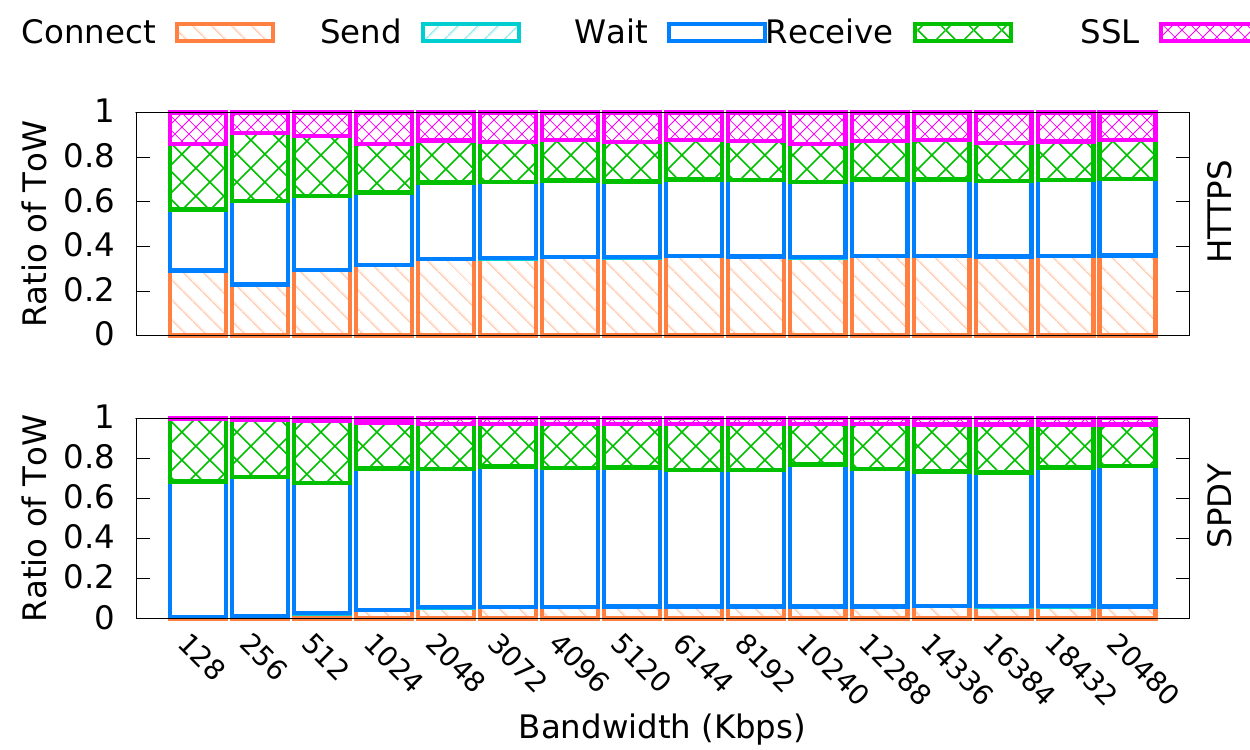}
  \caption{HAR Times for different Bandwidth values (YouTube, RTT=150ms, PLR=0\%).}
\label{fig:plot-HARtimesBars-youtube}
\end{figure}

While this explains SPDY's superior performance at low bandwidths, it does not explain its poor performance as capacities increase. As soon as bandwidth becomes sufficient to avoid the increased congestion caused by HTTPS, the benefits of SPDY begin to diminish.  This is particularly the case for websites with fewer resources, like Twitter. 
To understand this, we breakdown the operations performed by SPDY and HTTPS. Figure \ref{fig:plot-HARtimesBars-youtube} presents the results for YouTube as an example. Again, the two protocols have very different constitutions. HTTPS spends a large proportion of its time in the connect phase, setting up TCP and SSL. In contrast, SPDY spends the bulk of its time in the wait phase. Deep inspection reveals streams blocking until the connection is free to transmit. 
In line with our previous findings, this highlights that SPDY does not always do an effective job of multiplexing. Whereas, previously, this was caused by the complexity of the webpage, here it appears that high capacity transmission is also a challenge. 
Thus, as bandwidth increases, HTTPS can amortise the costs of TCP and SSL setup by exploiting the higher raw throughput afforded by opening parallel TCP sockets. We also observe that this situation occurs particularly when dealing with larger resources (e.g.\ images in Twitter), as window size can be scaled up before each connection ends in HTTPS. In contrast, SPDY appears to struggle to fill TCP's pipe as the server waits for new requests for each object from the client. Indeed, the Wireshark traces show TCP throughput reductions between 2\% and 10\% in the case of SPDY due to this problem compared to that of the parallel HTTPS connections. It would therefore seem that SPDY's default use of a single TCP connection might be unwise in circumstances of high bandwidth.

\vspace{\vspacecut}

%%%%%%%%%%%%%%%%%%%%%%%%%%%%%%%%%%%%%%%%%%%%%%
\subsection{Packet Loss Ratio}
\label{sec:results:network:loss}

\begin{figure}[tbh!]
  \centering
  \includegraphics[width=80mm]{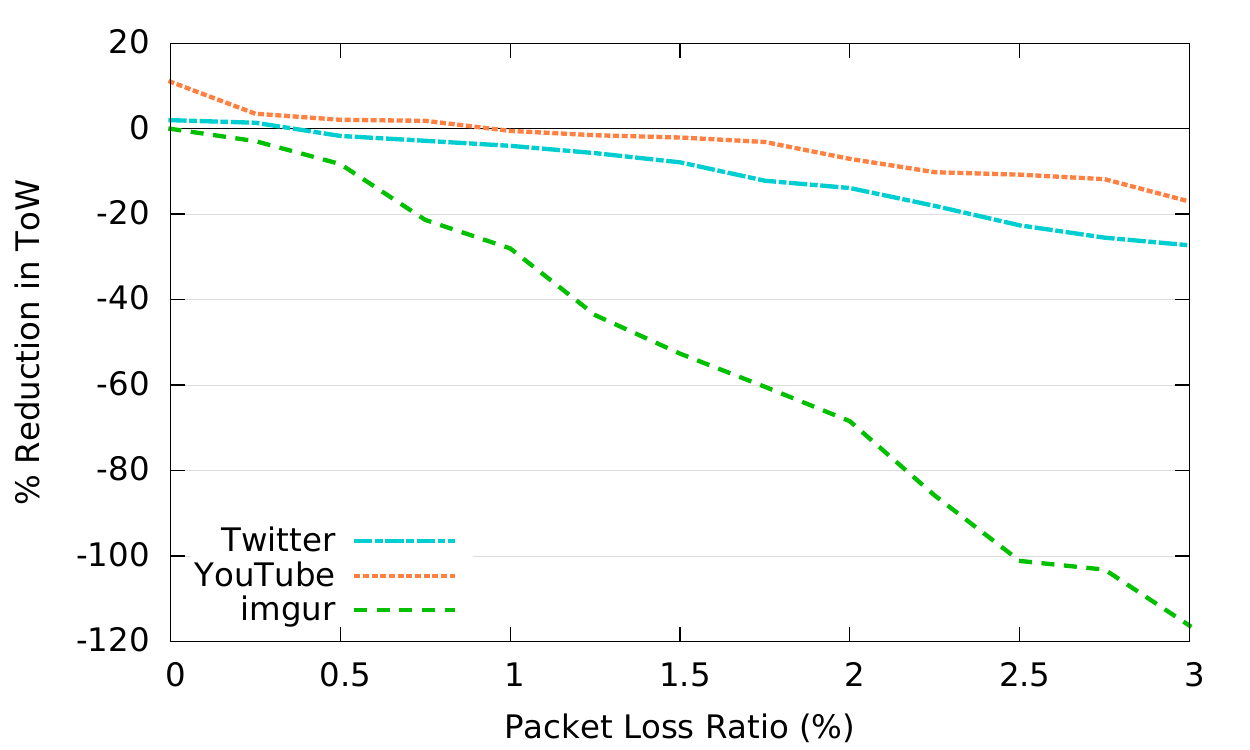}
  \caption{Effect of Packet Loss (RTT=150ms, BW=1Mbps).}
\label{fig:emu-loss-bw1-rtt150}
\end{figure}

Finally, we inspect the impact of packet loss on SPDY's performance. We fix RTT at 150ms and BW at 1Mbps, varying packet loss using the Linux kernel firewall with a stochastic proportional packet processing rule between 0 and 3\%\footnote{Packet loss in US mobile networks is reported to be as low as $\approx$ 0.2\% \cite{chen2012characterizing} and as high as 1.9\% \cite{heikkinen2012comparison}, but is considerably higher in other countries; e.g.\ 2-3\% in many European countries and $\geq$ 3\% in China, Russia and several South American states \cite{heikkinen2012comparison}. 
It is also quite high for WiFi \cite{Sheth2007loss}. We therefore consider 0--3\% to be an appropriate parameter range.}. Figure \ref{fig:emu-loss-bw1-rtt150} presents the results.

\vspace{\vspacecut}

Immediately, we see that SPDY is far more adversely affected by packet loss than HTTPS is. This has been anticipated in other work \cite{Thomas2012spdying} but never before tested. It is also contrary to what has been reported in the SPDY white paper \cite{SPDYWhitepaper}, which states that SPDY is better able to deal with loss. The authors suggest because SPDY sends fewer packets, the negative effect of TCP backoff is mitigated. We find that SPDY does, indeed, send fewer packets (upto 49\% less due to TCP connection reuse). However, SPDY's multiplexed connections persist far longer compared to HTTPS. Thus, a lost packet in a SPDY connection has a more profound setback on the long term TCP throughput than it would in any of HTTPS' ephemeral connections, the vast majority of which do not last beyond the TCP slow start phase \cite{Sun2011cdns}.  Furthermore, packet loss in SPDY affects all following requests and responses that are multiplexed over the same TCP connection. In contrast, a packet loss in one of the parallel HTTPS connections would not affect the other connections, neither concurrent nor subsequent (assuming HTTP pipelining is not used, which is commonly the case). In essence, HTTPS `spreads the risk' across multiple TCP connections. On average, we found that SPDY's throughput is affected by packet loss up to 7 times more than HTTPS (all experiments were performed using the default Linux CUBIC congestion avoidance algorithm).

It is also important to note that the probability of packet loss is higher in SPDY. According to \cite{Flach2013latency}, the probability of experiencing loss increases in proportion to the position of the packet in a burst chain. Hence, the chance of experiencing a packet tail drop is much higher for longer lived connections such as SPDY's. 
Thus, not only does SPDY react badly to packet loss, the chance of it experiencing loss is also higher. This is effectively highlighted in  Figure \ref{fig:emu-loss-bw1-rtt150}; imgur, which has the longest transfer time (by far), exhibits extremely poor performance under packet loss.

Finally, these results indicate that SPDY may not perform that well in mobile settings, one of its key target environments \cite{SPDY-perf-mobile}. Whilst both SPDY's high delay and low bandwidth support is desirable in this environment, the benefits can be undone by relatively low levels of packet loss (e.g.\ 0.5\%).

%%%%%%%%%%%%%%%%%%%%%%%%%%%%%%%%%%%%%%%%%%%%%%
\section{Effect of Infrastructural Decisions}
\label{sec:results:infra}
%%%%%%%%%%%%%%%%%%%%%%%%%%%%%%%%%%%%%%%%%%%%%%

\begin{figure}[htb!]
  \centering
  \includegraphics[width=80mm]{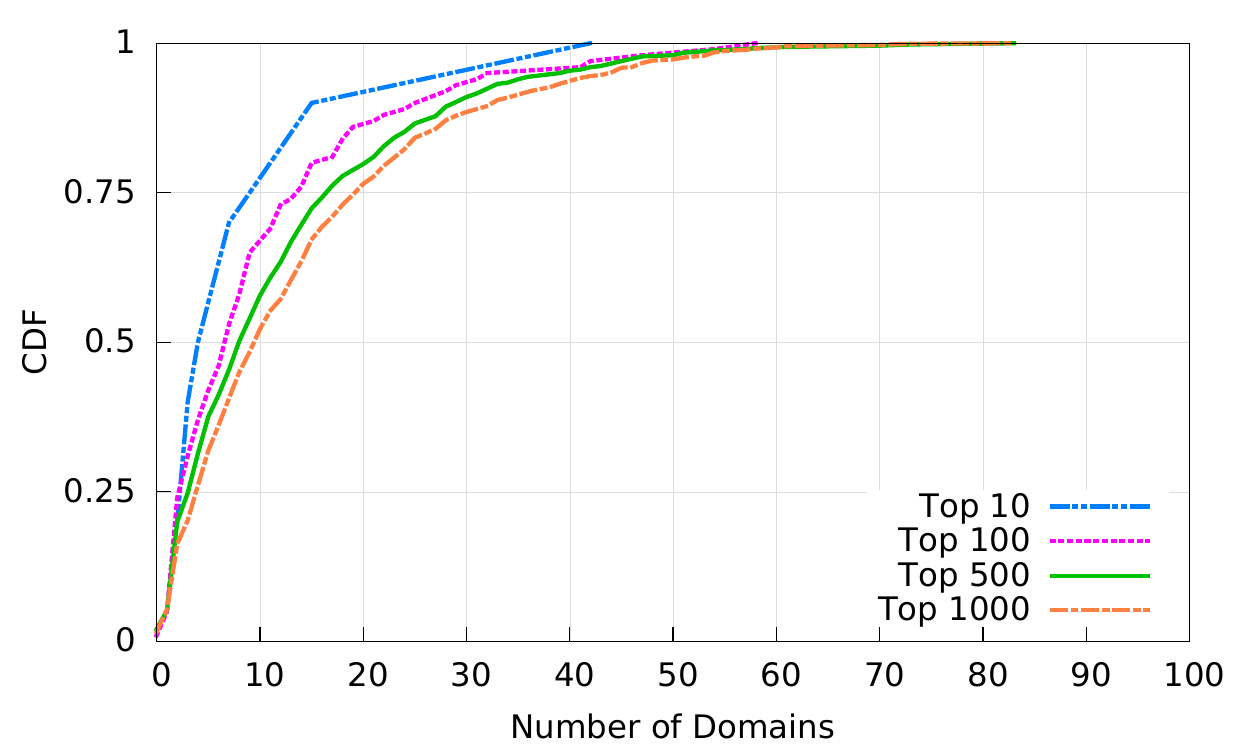}
  \caption{Number of Alexa Websites Resource Domains}
\label{fig:PageResourceDomainCount}
\end{figure}

The previous section has investigated the performance of SPDY under different network conditions 
between a single client and server. However, our original crawling of the Alexa Top 10k highlighted a tendency for providers to implement a practice known as \emph{domain sharding}. This is the process of distributing page resources across multiple domains (servers), allowing browsers to open more parallel connections to download page resources. 
Figure \ref{fig:PageResourceDomainCount} presents a CDF of the number of shards we discovered. We find that apart from front-less websites (such as media.tumblr.com and akamaihd.net), all websites employ some degree of domain sharding. 
Here, we choose to deep dive into the practice of domain sharding to understand the implications of this infrastructural design choice on SPDY.

\subsection{Number of Shards}
\label{sec:results:infra:shards}

\begin{figure}[htb!]
  \centering
  \includegraphics[width=80mm]{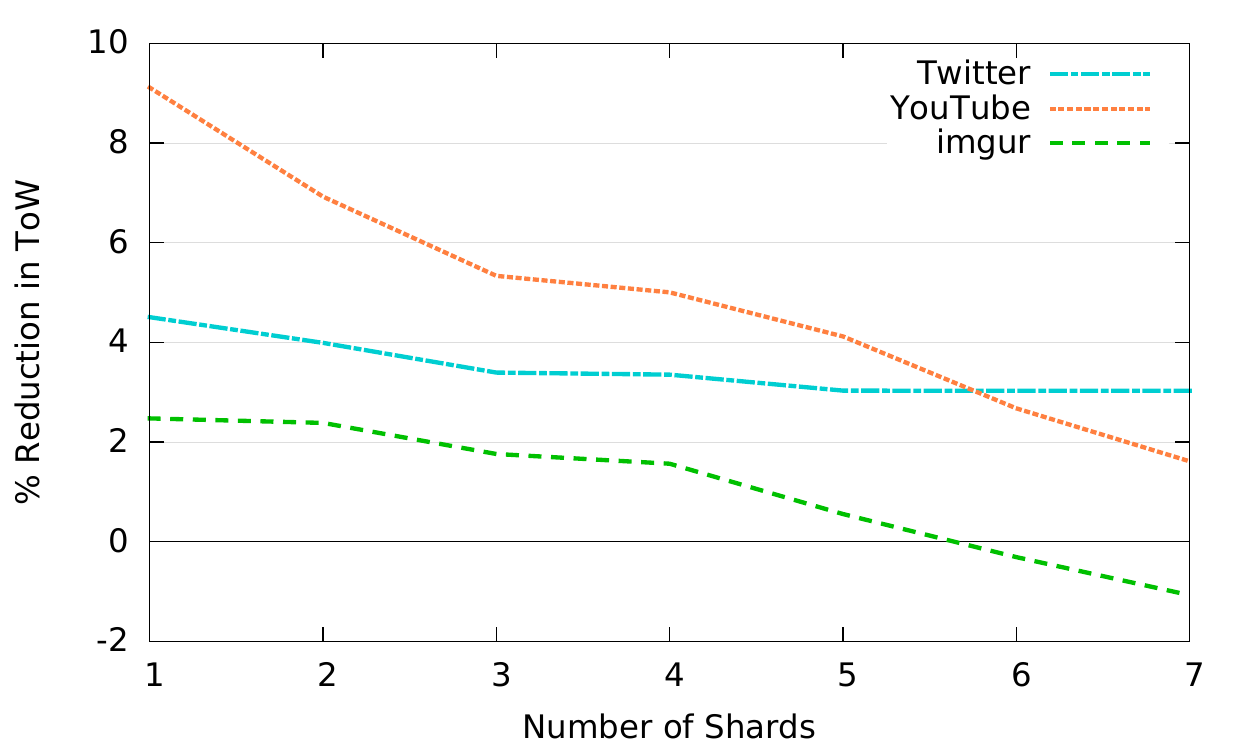}
  \caption{Effect of the Number of Shards}
\label{fig:plot-shards7}
\end{figure}

To inspect the impact of sharding, we recreate the earlier experimental setup but mirror the webpages across multiple servers, as occurs in real setups. We consider 7 shards, i.e.\ servers, an appropiate upper limit here as our measurements find that 70 of the top 100 Alexa websites have 7 or fewer shards. Each shard is configured as in Section \ref{sec:meth:setups:emu-net}. We then distribute the webpage resources across these servers. We perform retrievals 
using configurations between 1 and 7 shards, after adapting the HTML to reference shards in a round robin fashion. The client is configured with 1Mbps bandwidth, 150ms RTT and 0\% PLR. Figure \ref{fig:plot-shards7} presents the results of 100 runs at each configuration.

\vspace{\vspacecut}

We first note that sharding distinctly decreases SPDY's gain for YouTube and imgur. As the number of shards increases, so does the maximum number of parallel HTTPS connections. SPDY, too, is forced into creating multiple parallel TCP connections (one to each server). Hence, \emph{both} protocols are allowed to capitalise on increased parallelism. However, the benefits achieved by HTTPS outweigh those of SPDY as the former gains 6 new TCP connections per shard, a large performance boost that, in essence, offers SPDY-like multiplexing. This, therefore, reduces the overall improvement offered by SPDY. Another ramification of sharding evident from the examples of YouTube and imgur, is that as SPDY opens more connections, it multiplexes fewer streams per connection. This diminishes the returns of multiplexing which is SPDY's main competitive advantage over HTTPS. This suggests, based on our findings in Section \ref{sec:results:network:bw}, that increasing the servers' IW would give SPDY an advantage and the potential to tip the balance in its favour. 

The case of Twitter provides a different insight. Here, fairly steady results are achieved across all sharding levels. SPDY gains marginal improvements over HTTPS by reducing the number of round trips, which is dictated by the number of resources in a page. For such a page with only 7 resources, SPDY saves between one and two round trips at 1 shard (depending on whether all resources were requested together or at different times as the page is rendered). With more shards, the number of round trips that SPDY potentially saves is reduced to only one, if any, due to its reduced ability to multiplex. Whereas, in the case of HTTPS, more shards means fewer resources (and hence fewer parallel connections) per shard. This has the effect of gradually decreasing HTTPS' parallelism as the number of shards increase, hence allowing SPDY to continue to retain an edge.

In summary, we deduce that SPDY loses its performance gains as a website is sharded more. However, these negative results are not ubiquitous and vary remarkably depending on the number of page resources. This raises a few questions about SPDY deployment. Are the benefits enough for designers and admins to restructure their websites to reduce sharding? What about third party resources that cannot be consolidated, e.g.\ ads and social media widgets? Can SPDY be redesigned to multiplex across domains? Is proxy deployment \cite{Thomas2012spdying} rewarding and feasible as a temporary solution? The success of SPDY (and thereupon HTTP/2.0) is likely to be dependent on the answers to precisely these questions.

%%%%%%%%%%%%%%%%%%%%%%%%%%%%%%%%%%%%%%%%%%%%%%
\subsection{Number of Multiplexed Streams}
\label{sec:results:infra:mux}

So far, we have seen that sharding can create a significant challenge to SPDY's performance by forcing it into HTTP-like behaviour and by that limiting its ability to perform stream multiplexing. To further inspect this, we now directly study the impact of this multiplexing by artificially changing the maximum number of streams allowed per connection. This allows us to control exactly the degree of multiplexing afforded by SPDY. We vary this value from 1 to 100 (which is the default in Apache, as recommended by the SPDY draft \cite{SPDY-draft}) whilst mirroring the three websites on a single server. We perform these retrievals for a variety of RTTs. We choose to vary RTT because of the discovery that many of the bandwidth impacts are actually products of the inflated RTTs caused by queuing. Bandwidth is fixed at 1Mbps and PLR at 0\%.

\begin{figure}[tb!]
  \centering
  \subfloat[Twitter]{ \includegraphics[width=80mm]{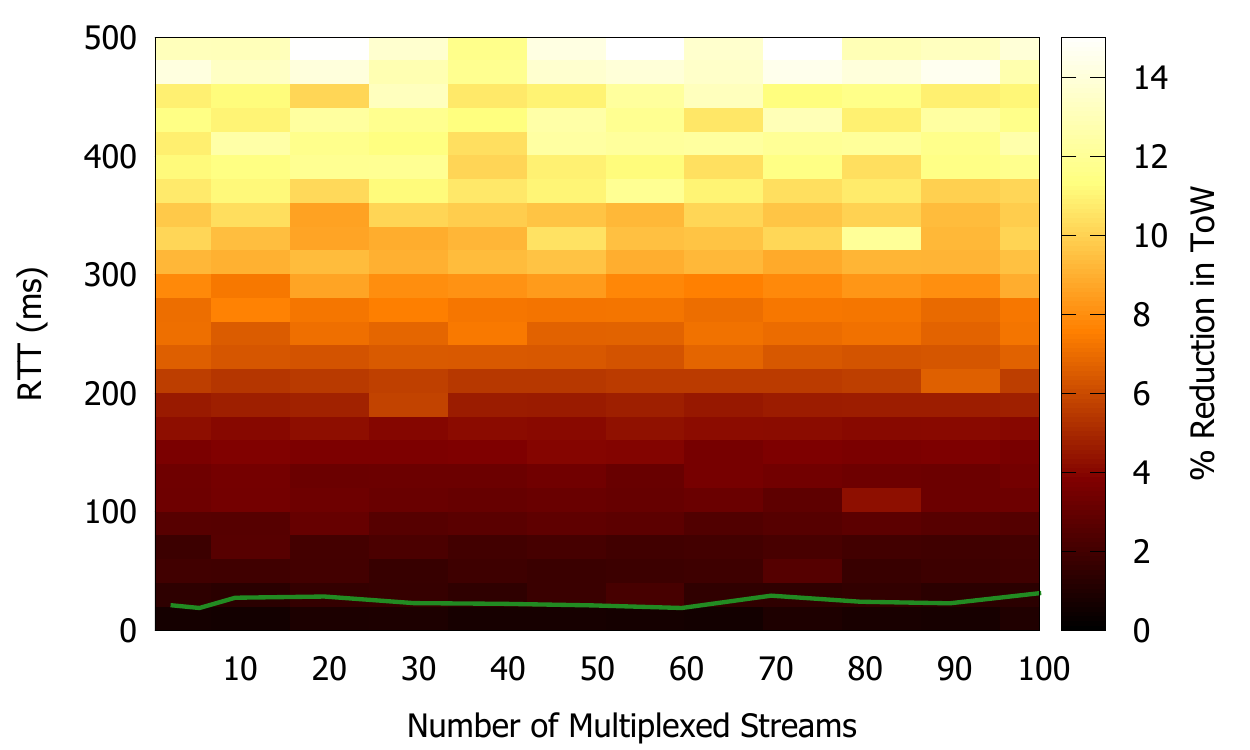} }    \\
  \subfloat[YouTube]{ \includegraphics[width=80mm]{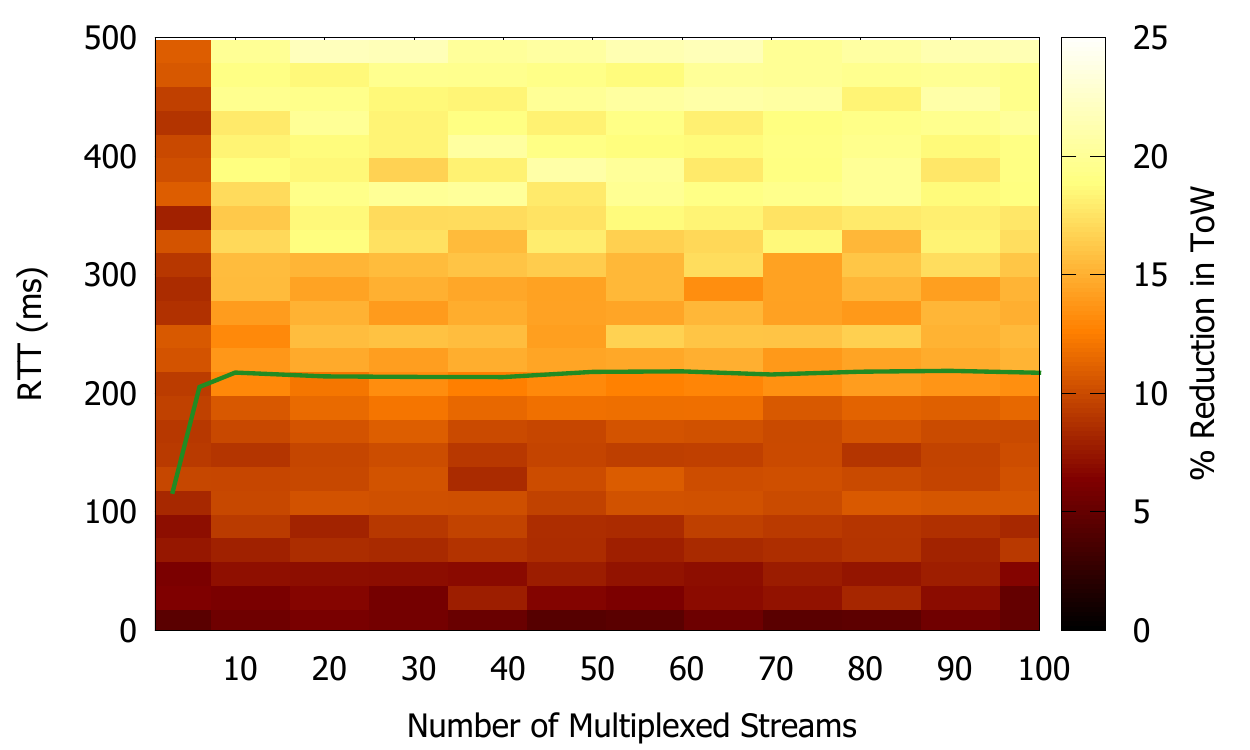} }    \\
  \subfloat[imgur] {  \includegraphics[width=80mm]{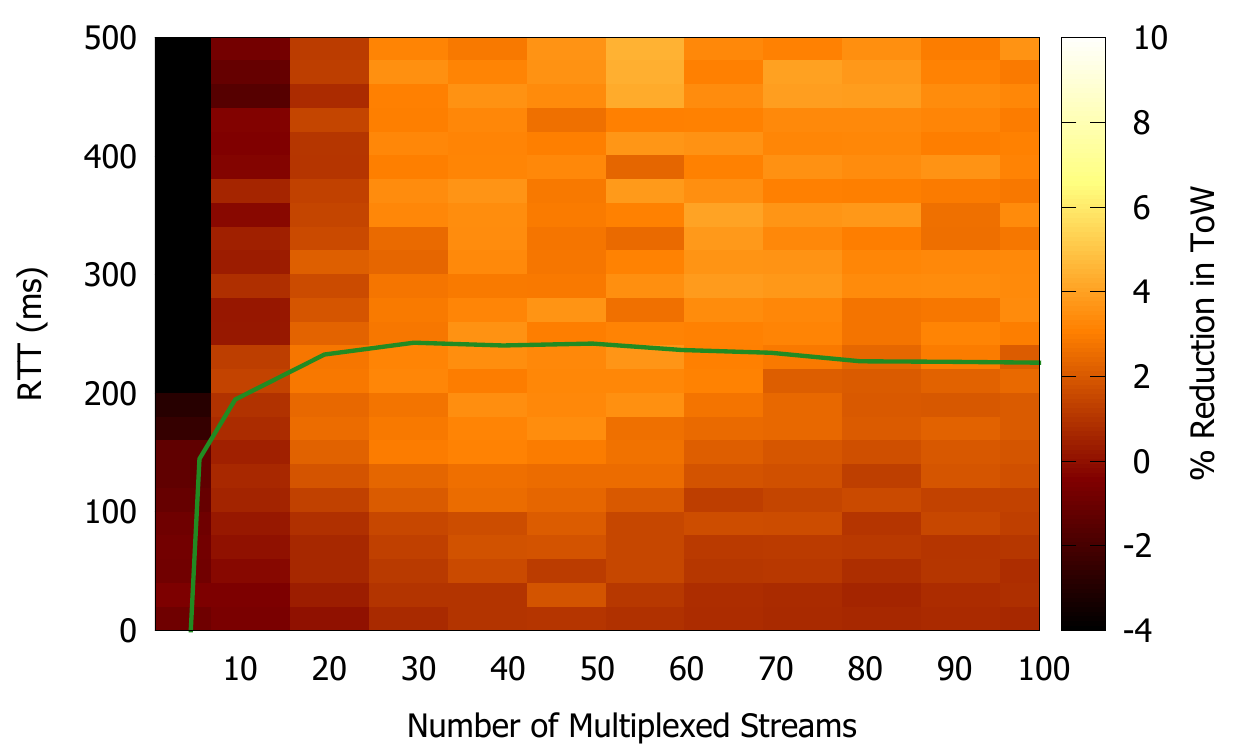} } \\
\caption{Effect of the Number of Multiplexed Streams per SPDY Connection over Varying RTTs.}
\label{fig:plot-heatmap-mux}
\end{figure}

In Figure \ref{fig:plot-heatmap-mux}, the average improvement in ToW (over HTTPS) for each multiplexing degree is displayed as a trend line, whilst the ToW reduction over different RTT values is shown as a heatmap to elicit more generalisable results. In all cases, SPDY's multiplexing has the potential to improve the ToW. For YouTube and imgur, we see a direct relationship between these benefits and the level of multiplexing afforded by SPDY. Here, these benefits plateau at 10 streams for YouTube and at 30 for imgur. In contrast, the results for Twitter 
remain relatively steady for all levels of multiplexing. 

To explore the different results for each page, we inspect the nature of their resources, as well as SPDY's recorded behaviour when accessing them. We confirm that these results are a product of the complexity of the webpages in terms of their resources. Twitter benefits little from increasing the multiplexing degree, as it only possesses 7 resources, i.e. no further benefits can be achieved when multiplexing beyond this level. The inverse case is found with YouTube (50 resources) and imgur (133 resources), which clearly can exploit multiplexing levels beyond 7 streams. Preventing this from happening has dire ramifications: when allowing SPDY to multiplex fewer than 6 streams for YouTube and imgur, it performs worse then HTTPS. This therefore confirms the negative impact that sharding will have on SPDY's deployment, where multiplexing capabilities could be severely undermined.  We found these observations to be true for even more complicated websites, e.g.\ the New York Times website (148 resources).

To better understand the relationship between performance and page complexity, we perform regression analysis to look at the multiplexing level ($m$) required to outperform HTTPS for a website with a given number of resources ($r$). This is done for all websites under test in addition to three other websites we experimented with. We find that $m \approx r/4$, with a very strong fit ($R^2=0.98537$, $p$-value$=8.0684\e{-5}$). This is not a robust model and is not intended to be so; it effectively highlights the impact that page type will have on SPDY's performance. Another interesting point here is that intuition would perhaps lead towards a $r/6$ relationship, due to the maximum number of parallel HTTPS connections. Instead, $m$ is found to be of greater value. We are not able to pinpoint the reasons behind this, but it could be attributed to SPDY's multiplexing overheads diagnosed in section \ref{sec:results:network}.

%%%%%%%%%%%%%%%%%%%%%%%%%%%%%%%%%%%%%%%%%%%%%%
\section{Conclusions \& Future Work}
\label{sec:Conclusion}
%%%%%%%%%%%%%%%%%%%%%%%%%%%%%%%%%%%%%%%%%%%%%%

SPDY provides a low-cost upgrade of HTTP, aiming to reduce page load times leading to improved user experience. To do this, it introduces a variety of new features, including stream multiplexing and header compression. Currently, the behaviour and performance of SPDY are quite poorly understood, exacerbated by the often conflicting results reported by various early stage studies. Our own live experiments confirmed these observations, highlighting SPDY's ability to both decrease and increase page load times.

We therefore turned our efforts to identifying the conditions under which SPDY thrives. We found that SPDY offers maximum improvement when operating in challenged environments, i.e. low bandwidth and high delay. We concluded that stream multiplexing is at the heart of SPDY's performance. This feature allows it to minimise the number of round trips required to fetch resources. It also facilitates more disciplined congestion control, which allows SPDY to outshine HTTP on low bandwidth links and promotes further network enhancements such as increasing TCP's IW. On the other hand, SPDY's multiplexed connections last much longer than HTTP's, which makes SPDY more susceptible to loss and the subsequent issues with TCP backoff.

We then investigated the impact of infrastructural decisions on SPDY's performance, namely the prevalent practice of domain sharding. We observed that SPDY's benefits are reduced in sharded environments where SPDY is prevented from maximising on multiplexing. We predict this could have palpable implications on website design and deployment strategies. Finally, we observed throughout our experiments that page type has huge influence on SPDY's performance: SPDY favours pages with more and larger resources, as opposed to pages with a very large number of small resources which induces perceptible multiplexing overheads.

So far, we have investigated only a subset of SPDY's overall parameter space and, thus, our future work intends to focus on expanding these experiments. This includes alternate network configurations, but also extends to inspecting other SPDY features (e.g. Server Push and Hint). A particularly important aspect of our future work is to formulate a better understanding of SPDY's behaviour in relation to the different page characteristics, which we have discovered to have a profound impact on performance. Finally, this work should feed into the wider discussion regarding HTTP/2.0, and the future of the web. 

%%%%%%%%%%%%%%%%%%%%%%%%%%%%%%%%%%%%%%%%%%%%%%
\section{Acknowledgments}
This work was partly supported by the NERC EVOp (NE/I002200/1) and EPSRC IU-ATC (EP/J016748/1) projects. We thank Dr. Rajiv Ramdhany for access to testbed resources, and Prof. Gordon S. Blair for his valuable feedback.

%%%%%%%%%%%%%%%%%%%%%%%%%%%%%%%%%%%%%%%%%%%%%%
\bibliographystyle{abbrv}
{\balance\bibliography{spdy,rfc}}

\begin{thebibliography}{10}

\bibitem{httparchive}
{HTTP Archive}.
\newblock \url{http://httparchive.org/}.

\bibitem{SPDYWhitepaper}
Spdy: An experimental protocol for a faster web.
\newblock \url{http://www.chromium.org/spdy/spdy-whitepaper}.

\bibitem{SPDY-draft}
M.~Belshe and R.~Peon.
\newblock {SPDY Protocol}.
\newblock \url{http://tools.ietf.org/html/draft-mbelshe-httpbis-spdy-00}, Feb
  2012.
\newblock IETF Network Working Group.

\bibitem{akamai2012q4soti}
D.~Belson.
\newblock Akamai state of the internet report. 5(4).
\newblock \url{http://www.akamai.com/stateoftheinternet}, 2012.

\bibitem{chrome-har-capturer}
A.~Cardaci.
\newblock {chrome-har-capturer}.
\newblock \url{https://github.com/cyrus-and/chrome-har-capturer}.

\bibitem{chen2012characterizing}
Y.-C. Chen, D.~Towsley, E.~M. Nahum, R.~J. Gibbens, and Y.-s. Lim.
\newblock Characterizing {4G} and {3G} networks: Supporting mobility with
  multi-path {TCP}.
\newblock Technical report, UMass Amherst Technical Report: UM-CS-2012-022,
  2012.

\bibitem{rfc6928}
J.~Chu, N.~Dukkipati, Y.~Cheng, and M.~Mathis.
\newblock {Increasing TCP's Initial Window}.
\newblock RFC 6928 (Experimental), Apr 2013.

\bibitem{htb}
M.~Devera.
\newblock Hierachical token bucket theory.
\newblock \url{http://luxik.cdi.cz/~devik/qos/htb/manual/theory.htm}, May 2002.

\bibitem{Dischinger2007broadband}
M.~Dischinger, A.~Haeberlen, K.~P. Gummadi, and S.~Saroiu.
\newblock Characterizing residential broadband networks.
\newblock In {\em Proc. IMC}, pages 43--56. ACM, 2007.

\bibitem{Elkhatib11}
Y.~Elkhatib.
\newblock {\em {Monitoring, Analysing and Predicting Network Performance in
  Grids}}.
\newblock PhD thesis, Lancaster University, Lancaster, UK, September 2011.

\bibitem{rfc2616}
R.~Fielding, J.~Gettys, J.~Mogul, H.~Frystyk, L.~Masinter, P.~Leach, and
  T.~Berners-Lee.
\newblock {Hypertext Transfer Protocol -- HTTP/1.1}.
\newblock RFC 2616 (Draft Standard), Jun 1999.
\newblock Updated by RFCs 2817, 5785, 6266, 6585.

\bibitem{Flach2013latency}
T.~Flach, N.~Dukkipati, A.~Terzis, B.~Raghavan, N.~Cardwell, Y.~Cheng, A.~Jain,
  S.~Hao, E.~Katz-Bassett, and R.~Govindan.
\newblock Reducing web latency: the virtue of gentle aggression.
\newblock In {\em Proc. SIGCOMM}, 2013.

\bibitem{Gettys2011Bufferbloat}
J.~Gettys and K.~Nichols.
\newblock Bufferbloat: Dark buffers in the internet.
\newblock {\em Queue}, 9(11):40:40--40:54, Nov 2011.

\bibitem{mod-spdy}
{Google SPDY project}.
\newblock {mod-spdy: Apache SPDY module}.
\newblock \url{http://code.google.com/p/mod-spdy/}.

\bibitem{httpbisTicket131}
H.~W. Group.
\newblock Ticket \#131: increase connection limit.
\newblock \url{http://trac.tools.ietf.org/wg/httpbis/trac/ticket/131}, Sep
  2008.

\bibitem{heikkinen2012comparison}
M.~V. Heikkinen and A.~W. Berger.
\newblock Comparison of user traffic characteristics on mobile-access versus
  fixed-access networks.
\newblock In {\em Proc. PAM}, pages 32--41. Springer, 2012.

\bibitem{hemminger2005netem}
S.~Hemminger.
\newblock Network emulation with {NetEm}.
\newblock In {\em Proc. of Linux Conf. Australia}, pages 18--23. Citeseer,
  2005.

\bibitem{Kaune2009modelling}
S.~Kaune, K.~Pussep, C.~Leng, A.~Kovacevic, G.~Tyson, and R.~Steinmetz.
\newblock Modelling the internet delay space based on geographical locations.
\newblock In {\em Proc. Conf. Parallel, Distributed and Network-based
  Processing}, pages 301--310, Feb 2009.

\bibitem{Kohavi2007experiments}
R.~Kohavi and R.~Longbotham.
\newblock Online experiments: Lessons learned.
\newblock {\em IEEE Computer Magazine}, 40(9):103--105, 2007.

\bibitem{Li2007measuring}
F.~Li, M.~Li, R.~Lu, H.~Wu, M.~Claypool, and R.~Kinicki.
\newblock Measuring queue capacities of {IEEE} 802.11 wireless access points.
\newblock In {\em Proc. BroadNets}, pages 846--853, Sep 2007.

\bibitem{nah2004study}
F.~F.-H. Nah.
\newblock A study on tolerable waiting time: how long are web users willing to
  wait?
\newblock {\em Behaviour \& Information Technology}, 23(3):153--163, 2004.

\bibitem{Padhye2012TR}
J.~Padhye and H.~F. Nielsen.
\newblock A comparison of {SPDY} and {HTTP} performance.
\newblock Microsoft Technical Report MSR-TR-2012-102, 2012.

\bibitem{NotAsSPDY}
G.~Podjarny.
\newblock Not as spdy as you thought.
\newblock \url{http://www.guypo.com/technical/not-as-spdy-as-you-thought/},
  2012.

\bibitem{Severance2012javascript}
C.~Severance.
\newblock {JavaScript}: Designing a language in 10 days.
\newblock {\em IEEE Computer Magazine}, 45(2):7--8, 2012.

\bibitem{Sheth2007loss}
A.~Sheth, S.~Nedevschi, R.~Patra, S.~Surana, E.~Brewer, and L.~Subramanian.
\newblock Packet loss characterization in {WiFi}-based long distance networks.
\newblock In {\em Proc. INFOCOM}, pages 312--320, 2007.

\bibitem{skadberg2004visitors}
Y.~X. Skadberg and J.~R. Kimmel.
\newblock Visitors' flow experience while browsing a web site: its measurement,
  contributing factors and consequences.
\newblock {\em Computers in human behavior}, 20(3):403--422, 2004.

\bibitem{Sun2011cdns}
P.~Sun, M.~Yu, M.~J. Freedman, and J.~Rexford.
\newblock Identifying performance bottlenecks in {CDNs} through {TCP}-level
  monitoring.
\newblock In {\em SIGCOMM workshop on Measurements Up the Stack}, pages 49--54.
  ACM, 2011.

\bibitem{Sundaresan2011broadband}
S.~Sundaresan, W.~de~Donato, N.~Feamster, R.~Teixeira, S.~Crawford, and
  A.~Pescap\`{e}.
\newblock Broadband internet performance: a view from the gateway.
\newblock In {\em Proc. SIGCOMM}, pages 134--145. ACM, 2011.

\bibitem{Thomas2012spdying}
B.~Thomas, R.~Jurdak, and I.~Atkinson.
\newblock {SPDYing} up the web.
\newblock {\em Commun. of ACM}, 55(12):64--73, Dec 2012.

\bibitem{SM-draft}
R.~Trace, A.~Foresti, S.~Singhal, O.~Mazahir, H.~F. Nielsen, B.~Raymor, R.~Rao,
  and G.~Montenegro.
\newblock {HTTP Speed+Mobility}.
\newblock
  \url{http://tools.ietf.org/html/draft-montenegro-httpbis-speed-mobility-02},
  Jun 2012.
\newblock IETF Network Working Group.

\bibitem{SPDY-perf-mobile}
M.~Welsh, B.~Greenstein, and M.~Piatek.
\newblock {SPDY Performance on Mobile Networks}.
\newblock \url{https://developers.google.com/speed/articles/spdy-for-mobile},
  Apr 2012.
\newblock Google `Make the Web Faster' project.

\bibitem{whiteIEFT}
G.~White, J.~Mule, and D.~Rice.
\newblock {Analysis of SPDY and TCP Initcwnd}.
\newblock
  \url{http://tools.ietf.org/html/draft-white-httpbis-spdy-analysis-00}, Jul
  2012.
\newblock IETF Network Working Group.

\end{thebibliography}

\end{document}